Hindawi

*Research Article*

# A New Weighting Scheme in Weighted Markov Model for Predicting the Probability of Drought Episodes

**Zulfiqar Ali** ,[1] **Ijaz Hussain** ,[1] **Muhammad Faisal** ,[2,3] **Ibrahim M. Almanjahie,**[4] **Muhammad Ismail** ,[5] **Maqsood Ahmad,**[6] **and Ishfaq Ahmad** [7,4]

[1]*Department of Statistics, Quaid-i-Azam University, Islamabad, Pakistan*
[2]*Faculty of Health Studies, University of Bradford, BD7 1DP Bradford, UK*
[3]*Bradford Institute for Health Research, Bradford Teaching Hospitals NHS Foundation Trust, Bradford, UK*
[4]*Department of Mathematics, College of Science, King Khalid University, Abha 61413, Saudi Arabia*
[5]*Department of Statistics, COMSATS University Islamabad, Lahore Campus, Pakistan*
[6]*Department of Mathematics, COMSATS University Islamabad, Lahore Campus, Pakistan*
[7]*Department of Mathematics and Statistics, Faculty of Basic and Basic Sciences, International Islamic University, 44000 Islamabad, Pakistan*

Correspondence should be addressed to Zulfiqar Ali; zulfiqarali@stat.qau.edu.pk

Drought is a complex stochastic natural hazard caused by prolonged shortage of rainfall. Several environmental factors are involved in determining drought classes at the specific monitoring station. Therefore, efficient sequence processing techniques are required to explore and predict the periodic information about the various episodes of drought classes. In this study, we proposed a new weighting scheme to predict the probability of various drought classes under Weighted Markov Chain (WMC) model. We provide a standardized scheme of weights for ordinal sequences of drought classifications by normalizing squared weighted Cohen Kappa. Illustrations of the proposed scheme are given by including temporal ordinal data on drought classes determined by the standardized precipitation temperature index (SPTI). Experimental results show that the proposed weighting scheme for WMC model is sufficiently flexible to address actual changes in drought classifications by restructuring the transient behavior of a Markov chain. In summary, this paper proposes a new weighting scheme to improve the accuracy of the WMC, specifically in the field of hydrology.

## 1. Introduction

Drought, a highest ranked natural hazard, is the main source of severe destructive effects on the planet [1]. In particular, its sustained consequences lead to sterilization of agricultural land and initiation of diseases. Factors associated with higher risk of drought are long duration of comparatively less rainfall, high rate of evapotranspiration, low relative humidity, high temperature, and high wind speed [2]. Moreover, many other environmental and ecological factors are also responsible for the recurrent occurrences of drought hazard. However, drought intensity, duration, and severity may vary from region to region. In recent decades, almost all the developing countries are facing water shortage due to continued expansion in agriculture, industrial, and energy sectors. Consequently, a perpetual increase in difference between water demand and renewable freshwater resources will lead to major social and economic issues [3].

To avoid and overcome the adverse effects of drought hazard, numerous studies have been conducted in various climatic regions. Several studies have proposed drought monitoring tools and forecasting methods to predict and quantify the risk associated with the recurrent occurrences of drought. Drought indices are the most frequently used ones around the world among various available tools due to their simplicity in structure, robustness, and popularity.

A list of drought indices along with its variable requirement is available in the literature [4]. Beside drought monitoring tools, a number of drought forecasting methods have been developed to assess the risk of future drought conditions [5, 6]. However, the tendency of using stochastic process to model uncertain phenomena is rapidly increasing, such as Markov chain is a promising approach to model dynamic activities [7]. A Markov chain is a discrete time stochastic process, which has the property that the future state of the process is independent from the past state, given the present state [8]. Furthermore, Markov chain models can be useful for forecasting future drought classes due to their multifaceted nature to enumerate uncertainties associated with these hydrometeorological variables [9–11]. However, it is difficult to adjust the transition probability matrix of Markov chain for the precise forecasting of succeeding events at short time scale.

To handle this structural issue, several studies used weights in Markov chain models to improve model accuracy and precision [12–16]. However, the selection of weights is purely a subjective approach. The assignment of these weights depends on the type of data or purely subjective. Dynamic weighting schemes are proposed to adjust Markov chain model [17–22]. Furthermore, an error and trial methodology can be used to increase the performance of the weighted Markov chain [23].

In this study, we aimed to develop a new weighting method to address the structural difficulties in traditional Markov chain while forecasting short-term drought states. We use the standardized precipitation temperature index (SPTI) for drought classification. As SPTI produces drought severity level (highest to lowest), therefore, prediction of highest severity to lowest severity and vice versa is questionable in the traditional Markov chain setting. The rationale of this study is to use interrater reliability measures of association as weights for accurate and precise forecasting of drought classification states under Weighted Markov Chain (WMC) setting. We use our proposed weighting scheme for one-month ahead forecasting of drought states of five meteorological stations in Pakistan.

## 2. Methodology

Nowadays, drought indices play a substantial role in determining drought classes. However, the methodological configuration of each index solely depends on the availability of climatic data and its prior historical data. In the current scenario, the uses and application of multiscalar drought indices, being more flexible and having the characteristics to determine drought at various time scales, are very common. In this paper, monthly drought classifications determined by the multiscalar drought indices are assumed to follow first-order Markov chain. To predict the probability of one-month ahead drought classes, a new weighting scheme to determine the weights for WMC model is provided by aligning the role of autocorrelation and Cohen Kappa measure.

The detail on the methodological structure of the proposed model weighting schemes is explained in the subsequent subsection.

### 2.1. The Multiscalar Drought Indices.
There are several procedures to report drought severity using multiscalar drought index. McKee et al. [24] developed SPI drought index, which is based on long-term precipitation record to quantify the precipitation scarcity for different time scales. One of the major advantages of using SPI index is that it can be used to monitor drought for various time scales.

Vicente-Serrano et al. [25] developed a new multiscalar drought index: the standardized precipitation evapotranspiration index (SPEI). In SPEI, the water balance model based on the difference between precipitation and potential evapotranspiration (PET) is used with similar estimation procedure of SPI. One major advantage of SPEI over SPI is inclusion of the effect of evaporation in rainfall data to characterize the regions under study.

Following the same methodology of SPI and SPEI, Ali et al. [26] developed standardized precipitation temperature index (SPTI) drought index to capture drought characterization in both cold and hot climate regions. However, appropriate selection of drought indices also depends on the availability of data and the nature of climatic zone. In this study, we employed SPTI methodology to characterize monthly historical behavior of drought hazard. Stepwise mathematical procedure of SPTI methodology is discussed in [27].

In this paper, we followed the classification criteria provided by McKee et al. [24] and Vicente-Serrano et al. [25]. Table 1 shows the ranges of SPTI values associated with drought classes.

### 2.2. Drought Classes as a Markov Chain.
A discrete Markov chain is a random process that describes a sequence of events from a set of finite possible states, whereas current event depends only on the preceding event. It has been commonly used to model uncertain events in various fields such as, engineering [28], economics [29], and physics [30]. In recent decades, the use of Markov chain is common in many applications to capture the behavior of drought classification states using multiscalar drought indices [31, 32]. For example, Mishra et al. [33] examined the distribution of drought interval time and mean drought interarrival time by their joint probability density functions and Markov chain approach. Shatanawi et al. [34] found that exact prediction of drought index values is impossible based on ARIMA model. However, early warning of drought can be detected from monthly Markov transition probabilities. Therefore, in drought modeling context, time series data SPTI drought classes for a single station can be considered as a sequence of ordinal drought classes. Consequently, historical series of drought classification states for a specified station can be embodied as a discrete Markov chain process. Here, we assume that any single class of drought in time series of SPTI depends on its previous class and then proceed to the construction of the transition probability matrix. It is just a statistical compliance that allows us to consider each drought class as a 1st order Markov chain. However, if each class depends on its previous two classes or nth order class, it is recommended to use 2nd order or higher order Markov model accordingly.

Table 1: Drought classification criteria of SPTI [27].

| SPTI values | Drought classes |
|---|---|
| ≥2 | Extremely wet (EW) |
| 1.5 to 1.99 | Severely wet (SW) |
| 1 to 1.49 | Moderate wet (MW) |
| 0.99 to −0.99 | Near normal (NN) |
| −1 to −1.49 | Moderate drought (MD) |
| −1.5 to −1.99 | Severe drought (SD) |
| ≤2 | Extreme drought (ED) |

### 2.3. The Proposed Weighting Scheme.

In this study, time series data on drought classes $Z_t$ determined by the SPTI drought index at one-month time scale are assumed as a series of ordinal correlated random variables. In this situation, instead of using traditional correlation between the quantitative series of SPTI at various time lags, interrater reliability measure Kappa is suggested to assess the relationship between stochastic processes of $Z_t$ and $Z_{t-1}$ of SPTI. Moreover, on the same rationale of using autocorrelation as weight, interrater reliability measure among the time series of ordinal classification of the SPTI index at various time lags can be considered in advance to predict the present drought class. In previous studies, Sengupta et al. [23] showed that the WMC method can make accurate predictions when the time series data exhibit its stochastic nature. Hereafter, in order to predict the occurrence of next drought class, the weighted average of interrater reliability coefficient at various lags as a weight is suggested to adjust the predictive probabilities. Therefore, instead of using correlation, the idea of using an interrater reliability measure for the ordinal classification is more rational.

The basic steps of the proposed weighting scheme in the WMC-based prediction of drought classes are as follows:

(1) Drought classification states and transition probability matrix:

Let $X_k$ be the time series of drought classes, where $X_k$ may assume the nominal droughts classes $c_1, c_2, c_3, \ldots, c_n$ depending on the classification criteria of the SPTI drought index. The transient behavior of each drought class can be represented by transition probability matrix in the following ways

$$
\begin{array}{cccc}
\text{Drought Classes} & c_1 & c_2 & \cdots & c_n,
\end{array}
$$
$$
\begin{array}{c}
c_1 \\ c_2 \\ \vdots \\ c_n
\end{array}
\begin{pmatrix}
p_{11} & p_{12} & \cdots & p_{1n} \\
p_{21} & p_{22} & \cdots & p_{2n} \\
\vdots & \vdots & \ddots & \vdots \\
p_{n1} & p_{n2} & \cdots & p_{nn}
\end{pmatrix}, \qquad (1)
$$

where $c_1$, $c_2$, $c_3$, $\ldots$, $c_n$ represent drought classes corresponding to their transient probabilities matrix.

(2) Construction of transition probability matrix:

In this step, we classify SPTI drought index with estimated one-month time scale according to the classification criteria (see Table 1).

Furthermore, let $Y_{ij}^t$ be the number of transitions from the state $S_i$ to $S_j$ through $t$ steps in time series length of drought classes $X_k$ calculated from one-month time scale. Here, the transition probabilities for various time steps and various drought classes can be obtained by the following equation:

$$
P_{ij}^{(t)} = \frac{Y_{ij}^{(t)}}{Y_i}, \quad \text{For } i, j = 1, 2, \ldots, m, \qquad (2)
$$

where $t$ represents the order of the Markov chain. Here, the transition probability matrix for various existing drought classes at the previous $t$ time step is represented as

$$
P^{(t)} = \begin{bmatrix}
p_{11}^{(t)} & p_{12}^{(t)} & \cdot & \cdot & p_{1m}^{(t)} \\
p_{21}^{(t)} & p_{22}^{(t)} & \cdot & \cdot & p_{2m}^{(t)} \\
\cdot & \cdot & \cdot & \cdot & \cdot \\
\cdot & \cdot & \cdot & \cdot & \cdot \\
p_{m1}^{(t)} & p_{m2}^{(t)} & \cdot & \cdot & p_{mm}^{(t)}
\end{bmatrix}. \qquad (3)
$$

(3) Interrater reliability measures as weights:

Let $p_{ij}$ denote the proportion of drought classes $i$ and $j$ determined by the SPTI at lag $t-1$, where $i$ and $j$ represent time series on drought classes and $t$ represents time point. Then, the weighted Kappa at time $t$ lag as a measure of association for ordinal categorical sequence is defined as

$$
\widehat{U}_t = \frac{L_{ct} - cL_{ct-1}}{1 - cL_{ct-1}}, \qquad (4)
$$

where

$$
L_{ct} = \sum_{i=1}^{d-1} \sum_{i=2}^{d} w_{ij} p_{ij}, \qquad (5)
$$

and

$$
L_{ct-1} = \sum_{i=1}^{d-1} \sum_{i=2}^{d} w_{ij} p_i p_j, \qquad (6)
$$
$$
w_{ij} = (i - j)^2,
$$

where $w_{ij}$ is the squared weighted function in Cohen Kappa [35, 36] as suggested by Robieson [37] and King and Chinchilli [38] and $p_i$ and $p_j$ represent the marginal proportion that assigned to the drought classes $i$ and $j$. In this study, the formula of weighted Kappa is modified on the same rationale of autocorrelation. We used *irr* [39] R package to compute the values of weighted kappa at various time lags.

(4) Standardization of weights:

The weights for the weighted Markov chain model are computed by standardizing Kappa coefficient $\widehat{K}_{wt}$ computed by the following equation:

$$
W_t = \frac{\widehat{U}_{wt}}{\sum_{i=1}^{t} \widehat{U}_{wt}}. \qquad (7)
$$

At this moment, the time step for weighted Kappa is decided according to the steady state nature of the transition probability matrix. If transition probability matrix approaches their steady state at $s$ time point, then, we calculate weighted Kappa from 1 to $s$ steps accordingly.

(5) Prediction

In this step, we assume recent past states of drought classification states as an initial drought class and combine it with the row vectors of their corresponding transition probability matrix. Here, the state transition probability vectors can be expressed in the following form.

$$P_i^{(t)} = \left(P_{i1}^{(t)}, P_{i2}^{(t)}, P_{i3}^{(t)}, \ldots, P_{im}^{(t)}\right), \tag{8}$$

where $i$ represents the drought classes and $t$ represents the order of Markov chain.

Furthermore, those transition probability vectors that are in acceding power for all the previous candidate of drought classes are selected. The following $m$th order transition probability matrix shows how one can understand the above stated argument:

$$\begin{bmatrix} p\xi 1^{(1)} & p\xi 2^{(1)} & . & . & p\xi m^{(1)} \\ p\lambda 1^{(2)} & p\lambda 2^{(2)} & . & . & p\lambda m^{(2)} \\ . & . & . & . & . \\ . & . & . & . & . \\ p\rho 1^{(m)} & p\rho 2^{(m)} & . & . & p\lambda m^{(m)} \end{bmatrix}, \tag{9}$$

where $\zeta, \lambda, \rho \in S_i$, and $\zeta, \lambda, \rho \leq m$.

Here, $\zeta, \lambda$, and $\rho$ show the previous candidate drought classes and $m$ represents the order of Markov chain corresponding to each candidate drought classes. Finally, by Equation (10), we assigned the weights to each vector of candidate drought classes and get the predicted probabilities for each drought class. Those drought classes which have a maximum prediction probability $\max\{P_i \in S\}$ are then selected as predicted drought classes. Additionally, by iterating this algorithm, we can forecast $n$-time drought classes under the WMC framework.

$$P_i^* = \sum_{t=1}^{m} W_t P_i^t. \tag{10}$$

Amalgamation of the proposed weighting scheme with SPTI-based historical drought classification is validated based on five meteorological stations located in various regions of Pakistan. We used secondary data on monthly total precipitation and mean temperature data to illustrate WMC based on the prediction process. These datasets were obtained from the Karachi Data Processing Center through Pakistan Meteorological Department. A brief description of the study area and the application of the proposed framework for one-step ahead prediction of drought are provided in the following section.

## 3. Application

*3.1. Study Area.* Illustration of the proposed weighting scheme accomplished for the five meteorological stations of Pakistan includes Astor, Chilas, Cherat, Skardu, and Peshawar. Figure 1 shows the location map of the study area.

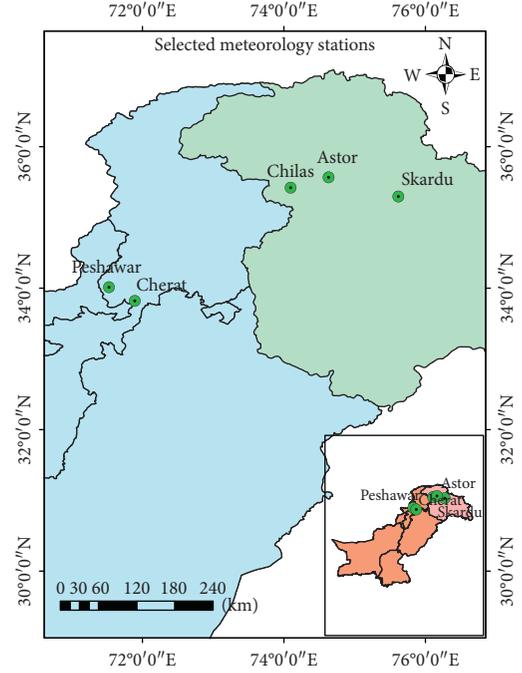

Figure 1: Map of selected locations of study area.

These stations have high variability in rainfall throughout the season. In each season, some of the stations are continuing to bear extremely vulnerable drought conditions. The drought has become a recurrent phenomenon in the country. In the recent decade, due to severe drought hazard, the economic system of the country was badly disturbed. Most parts of the country are arid to semiarid, with large spatial variability in the temperature, except the southern slopes of the Himalayas and the submountain region where the annual rainfall ranges from 760 mm to 2000 mm [40]. Pakistan has four well-marked seasons: cold, from November to February; premonsoon (hot), from March to mid of June; monsoon, from mid of June to mid of September; postmonsoon, from mid of September to October [41]. Summer season is extremely hot and the relative humidity ranges from 25% to 50%. The major part of Pakistan is arid to semiarid with large spatial variability in the temperature. In recent decades, several authors had been working to explore the geographical and hydrological importance of these stations. Awan [42] and Archer and Fowler [43] explored and inferred different climatic variables in terms of regression, spatial correlation, and temporal variation. Ahmad et al. [44] evaluated the significance of these mountainous areas that have a substantial potential in hydropower production and water resources. Ali et al. [26] compared the performance of SPTI with SPI and SPEI using time series data on precipitation and temperature on these stations. We applied our proposed model on long-term time series data monthly precipitations, minimum and maximum rainfall that was recorded during January 1955 to December 2017. The data were collected from Karachi Data Processing Center (KDPC) through Pakistan Meteorological Department (PMD). This dataset fulfills the WMO requirements, where errors, scrutiny, tabulation, and quality control are done by KDPC http://www.pmd.gov.pk/rmc/RMCK /Services_Climatology.html. The following two steps describe the detailed procedure for the estimation of SPTI.

The first step consists on the searching of appropriate probability distributions of DAI as suggested by Stagge et al. [45]. Therefore, the entire computational procedure of SPTI values is based on varying probability model for each station. Consequently, the estimation procedure consists of the searching of appropriate candidate distribution using Kolmogorov-Smirnov and Chi-Square, Anderson-Darling tests at the most commonly used level of significance = 0.5 by using *easyfit* software [46].

In the second step, several parameter estimation methods (method of moments, method of maximum likelihood estimation, and method of L-moments) are incorporated using R package *lmom* [47]. Probability distributions that have minimal value of the Akaike information criteria (AIC) is then standardized to obtain temporal values of each index accordingly.

Figures 2 and 3 represent the fitted probability distribution and their corresponding temporal values of SPTI for Astor and Cherat stations, respectively. The resulting values of SPTI are classified according to its classification criteria. To see the exploratory behavior of various drought classes, Figures 4 and 5 represent the cumulative frequencies of drought classes and its transition behavior of moving one drought class to another at Astor observatory. In this station, most of the month continued to bear near-normal weather conditions; however, as compared with other drought categories, extremely dry and severe dry drought classes are quite high.

*3.2. Results.* To test and infer the proposed framework of WMC-based prediction, we first compute transition probability matrix at various order form historical classifications of drought classes. In the current research, we use *Markovian chain* [48] R package to construct the transition probability matrices for all the stations. Secondly, weights at various lags are computed from ordinal categorical data on historical classification of drought classes by using Equations (4) and (7). To illustrate the steps associated with proposed framework, we provide the stepwise numerical results of Astor station. Therefore, Table 2 is especially prepared to show the value of Kappa associated with weights that are computed from the temporal classification of ordinal drought classes for Astor station. To predict future drought class, the original data on drought classes from June 2017 to December 2017 is arranged in chronological order. For example, June 2017 to December 2017, each month bears near normal situation. These drought classes are taken from the original classification of the SPTI drought index. To infer the probabilities of next drought classes, transition probability vector for each order is organized in matrix form. Table 3 shows the transition probability matrix in varying order with corresponding weights.

In the next step, according to the Equation (10), the weighted sums of the probabilities are computed for each drought class. In this numerical example, near normal drought class receives a maximum probability of occurrences ($P = 0.7146$) in January 2018. The actual drought category is also near normal. Hence, the method performs correct prediction. However, the next tentative class is moderately wet with a probability of 0.1146. Following Zhou et al. [16] and Zhang et al. [49], by taking predicted drought class as a reference category, the whole process may iterate to predict drought classes for March 2018 and so forth. Here, interrater reliability measure Kappa playing a role to adjust the long run convergence error in the Markov chain.

In line with Astor station, numerical investigations are carried out for Skardu, Chilas, Cherat, and Peshawar stations. A visual representation of one-step transition probability matrix can be seen in Figure 6, where the temporal profiles of drought classes in Peshawar and Cherat are explored. In these stations, each drought class has a high probability to transit in near-normal drought class. To adjust the temporal behavior of the Markov chain, it is assumed that by assigning appropriate weights to each drought classification, accurate forecasting of one or $n$ month ahead may conclude under the weighted Markov chain framework. Therefore, accuracy of the proposed model is assessed by cross-validation of the predicted drought class with original classification. Consequently, we left January 2018 for the validation phase of the proposed framework for all the stations.

Table 4 shows the values of squared Kappa with corresponding $P$ values in each time lag, its standardized values (weights), and the one-month ahead predicted probabilities of each drought class, where Chilas station has a high probability ($P = 0.4869$) of near-normal drought class on January 2018. In the same way, Cherat and Peshawar also have the same drought classification with probabilities 0.6948 and 0.6623, respectively. Since in these stations, the observed quantities of rainfall and the historical time series of rainfall quantities are quite high in December and January. Moreover, the original series of SPTI drought classes depicted wet classes for these two months in all stations. Contrary to the near normal drought class, Skardu observatory will bear severe drought class with a probability of 0.5003.

## 4. Discussion

This paper provides a new procedure to handle the prediction problem of ordinal categorical series of drought classes determined by SPTI drought index. For assessing the proposed method, we took five meteorological stations located in different climatology of Pakistan (see Figure 1). In this research, time series data of rainfall and temperature ranges from 1955 to 2017 recorded at monthly time scale were used to classify according to the classification criterion given in Ali et al. [26]. By assuming each temporal series of drought classes as a first-order Markov chain, the current research employed *Markovian chain* [48] R package to construct the transition probability matrices. According to the setting of WMC, the proposed weighted schemes were used to predict future drought condition of all the stations. We provide stepwise procedure for Astor station (see Table 2

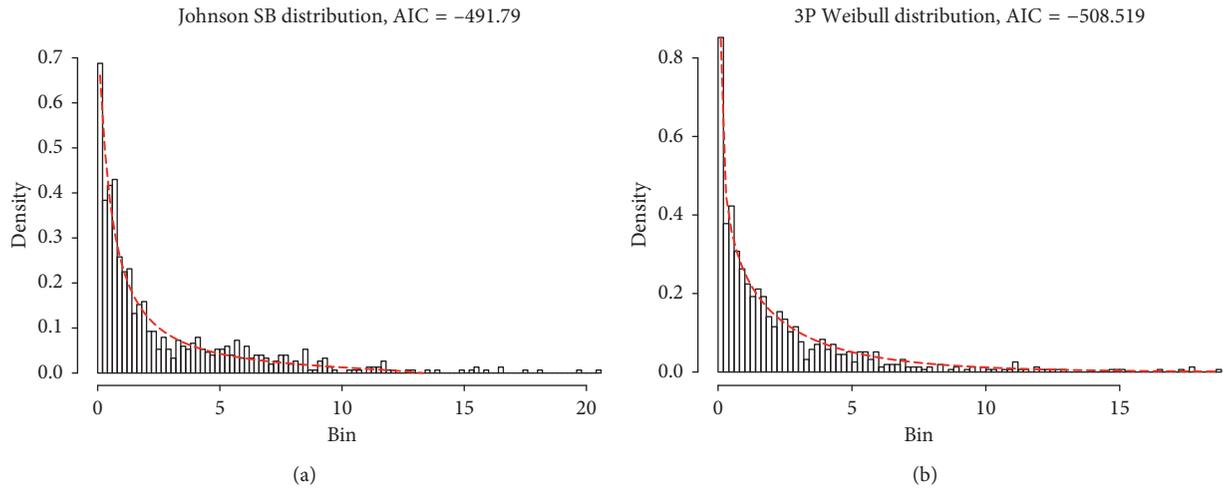

Figure 2: Probability distributions fitted for SPTI. (a) Astor. (b) Chilas.

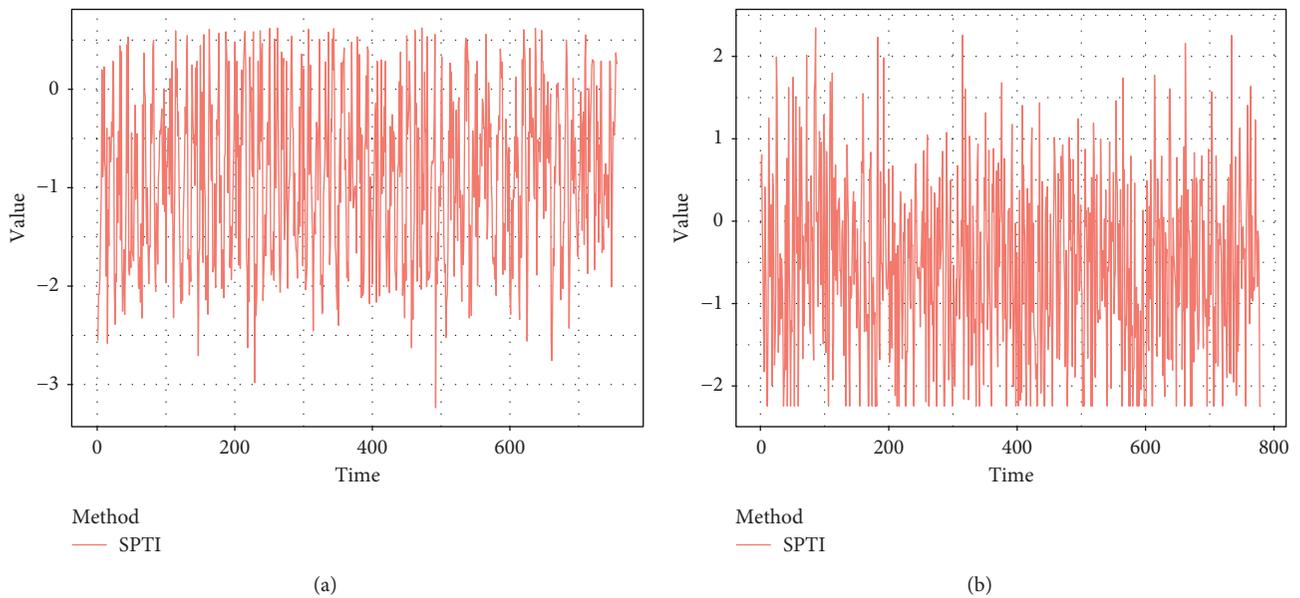

Figure 3: Temporal behavior of SPTI. (a) Astor. (b) Chilas.

and 3). However, prediction results are given for all the stations (see Table 4).

For assessing the consistency of the proposed method, we compared our prediction results with the steady state probabilities of each drought class. These long-term probabilities can also be viewed to cross validate the observed probabilities. It is observed that, in Chilas, Cherat, and Peshawar, the predicted drought classes are consistent with its long-term probabilities. However, in Skardu, a significant difference is observed, where long-term probability of near-normal drought class is 0.0563 and the predicted probability is 0.2483. This reflects the appropriateness of the proposed weighting scheme for ordinal classification of discrete stochastic process.

In summation, outcomes associated with this research show that the proposed weighting scheme may incorporate to adjust the structure of traditional Markov chain for short-term prediction. Additionally, our detailed analysis has also proved the suitability using interrater reliability instead of autocorrelation, as a weight in the WMC model. Therefore, trend from high to low accuracy can be controlled by adjusting the structural behavior of transition probability vectors from the proposed weighting scheme. However, the limitation of the proposed methods is not to consider the nonstationary behavior of Markov chain. Moreover, in the computations, the study assumed each Markov chain as first-order Markov process.

## 5. Conclusion

Prediction and forecasting play a very important role, especially in early warning situations. Consequently, accurate and precise techniques of drought forecasting may reduce their severe effect by making effective drought mitigation policies. In this article, the SPTI drought index being a more comprehensive drought monitoring procedure is used to classify historical monthly drought profile. Outcomes show that by introducing standardized squared weighted Kappa as a weight, the research suggests a new way to get adjusted prediction probabilities under WMC framework. Furthermore, it is observed that the advantage of *n* step forecasting can be achieved by just changing the transition probability vector and rearranging vectors of weights. Therefore, among numerous other studies and forecasting framework, the uniqueness of this research is to introduce ordinal measure of association at various lags in WMC-based prediction method. Consequently, by using SPTI or other multiscalar drought indices such as SPI and SPEI, where meteorological stations are characterized monthly ordinal drought classification, it is more reasonable to use ordinal measure of association, instead of correlation.

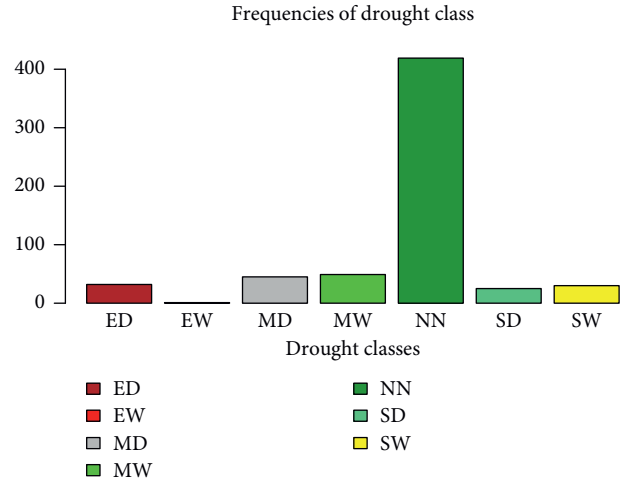

Figure 4: Frequencies of drought classes at Astor station.

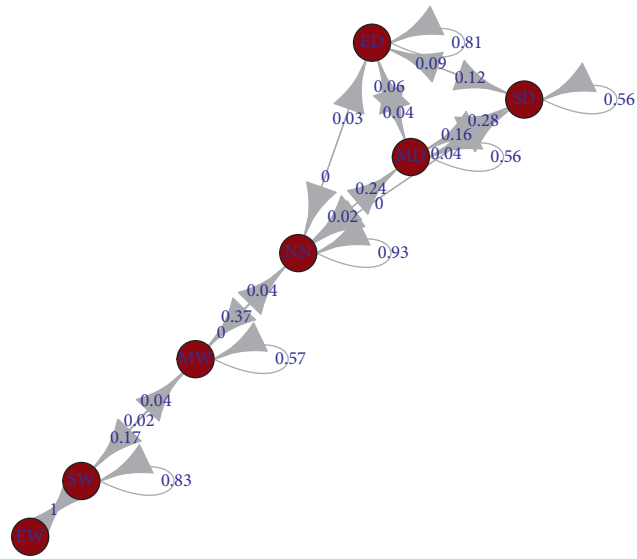

Figure 5: Transition behavior of drought classes at Astor station.

Table 2: Squared Kappa and weights at various time lags at Astor station.

| Step | 1 | 2 | 3 | 4 | 5 | 6 | 7 |
|------|------|------|------|------|------|------|------|
| Kappa | 0.812 | −0.0512 | 0.0382 | −0.0411 | 0.0083 | −0.0746 | 0.003 |
| Weights | 0.7895 | 0.0498 | 0.0371 | 0.04 | 0.0081 | 0.0725 | 0.0029 |

Table 3: Predicted probabilities of next drought classes at Astor station.

| Initial month | Class | Steps | Weights | ED | EW | MD | MW | NN | SD | SW |
|------|------|------|------|------|------|------|------|------|------|------|
| Jun | NN | $P^7$ | 0.00290 | 0.00000 | 0.05020 | 0.06700 | 0.13640 | 0.68420 | 0.01670 | 0.04550 |
| July | NN | $P^6$ | 0.07250 | 0.00240 | 0.06680 | 0.06440 | 0.13130 | 0.66830 | 0.01910 | 0.04770 |
| Aug | NN | $P^5$ | 0.00810 | 0.00240 | 0.05950 | 0.06670 | 0.12860 | 0.67860 | 0.01430 | 0.05000 |
| Sep | NN | $P^4$ | 0.04000 | 0.00000 | 0.05700 | 0.09260 | 0.11880 | 0.66510 | 0.01900 | 0.04750 |
| Oct | NN | $P^3$ | 0.03710 | 0.00240 | 0.06400 | 0.07110 | 0.11370 | 0.69670 | 0.01660 | 0.03550 |
| Nov | NN | $P^2$ | 0.04980 | 0.00240 | 0.05910 | 0.08750 | 0.10640 | 0.69270 | 0.01890 | 0.03310 |
| Dec | NN | $P^1$ | 0.78950 | 0.00240 | 0.03770 | 0.06840 | 0.11320 | 0.72410 | 0.01890 | 0.03540 |
| $P_i^* = \sum_{t=1}^{m} w_t P_i^t$ | | | | 0.00230 | 0.04290 | 0.07010 | 0.11460 | 0.71460 | 0.01880 | 0.03680 |

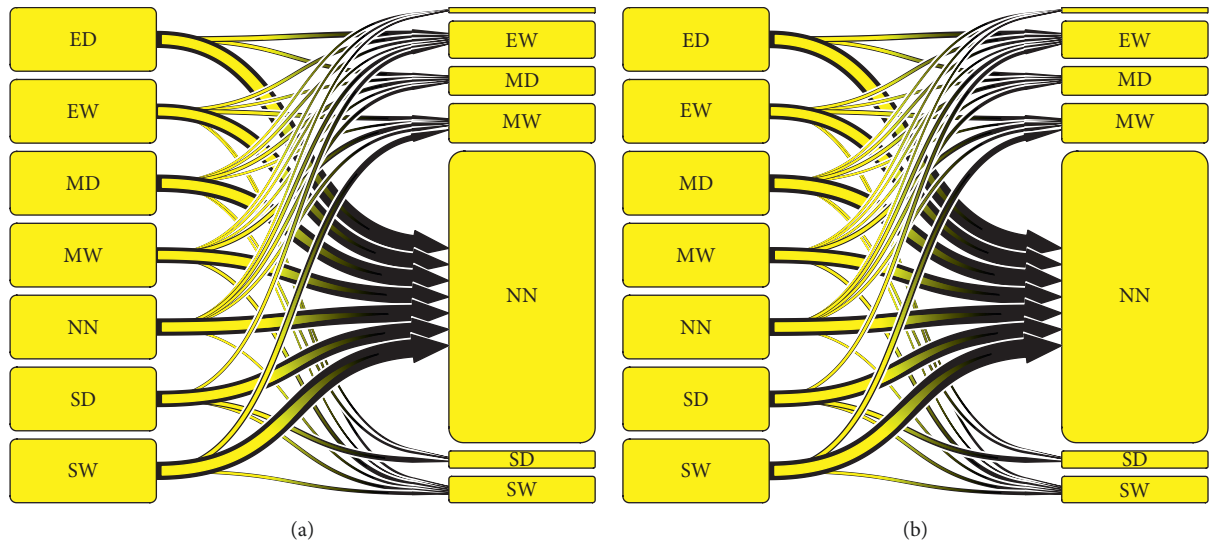

FIGURE 6: Visual representation of drought classes transitions. (a) Peshawar. (b) Cherat.

TABLE 4: One-month ahead probabilities under weighted Markov chain and Steady State behavior.

| Stations | Statistics | ED | EW | MD | MW | NN | SD | SW |
|---|---|---|---|---|---|---|---|---|
| Skardu | Kappa | 3.1794 | 0.9962 | −0.5078 | 0.2383 | −1.5348 | −2.0740 | Nil |
| | *P* value | 0.0015 | 0.3191 | 0.6116 | 0.8117 | 0.1248 | 0.0381 | Nil |
| | Weights | 0.3727 | 0.1168 | 0.0595 | 0.0279 | 0.1799 | 0.2431 | Nil |
| | IRMWMC | Nil | 0.0250 | 0.0702 | 0.1263 | 0.2483 | **0.5003** | 0.0299 |
| | Steady states | Nil | 0.0013 | 0.0034 | 0.2974 | 0.0563 | 0.6221 | 0.0195 |
| Chilas | Kappa | 1.6412 | 2.7045 | 0.5361 | 0.5355 | −1.2023 | 0.1986 | Nil |
| | *P* value | 0.1008 | 0.0068 | 0.5919 | 0.5923 | 0.2293 | 0.8426 | Nil |
| | Weights | 0.2407 | 0.3967 | 0.0786 | 0.0785 | 0.1763 | 0.0291 | Nil |
| | IRMWMC | Nil | 0.0004 | 0.1488 | 0.0445 | **0.4869** | 0.2616 | 0.0578 |
| | Steady states | Nil | 0.0025 | 0.1626 | 0.0560 | 0.4902 | 0.2646 | 0.0242 |
| Cherat | Kappa | 2.6746 | −0.7443 | 0.0175 | −0.0322 | −0.6303 | 0.0663 | 0.0949 |
| | *P* value | 0.0075 | 0.4567 | 0.9861 | 0.9743 | 0.5285 | 0.9471 | 0.9244 |
| | Weights | 0.6278 | 0.1747 | 0.0041 | 0.0076 | 0.1480 | 0.0156 | 0.0223 |
| | IRMWMC | 0.0241 | 0.0408 | 0.0162 | 0.1391 | **0.6948** | 0.0214 | 0.0637 |
| | Steady states | 0.0085 | 0.0938 | 0.0521 | 0.0948 | 0.6978 | 0.0128 | 0.0403 |
| Peshawar | Kappa | 1.6807 | −1.9287 | −0.0091 | −1.1911 | −1.9143 | −0.1484 | 0.6617 |
| | *P* value | 0.0928 | 0.0538 | 0.9928 | 0.2336 | 0.0556 | 0.8820 | 0.5082 |
| | Weights | 0.2231 | 0.2560 | 0.0012 | 0.1581 | 0.2541 | 0.0197 | 0.0878 |
| | IRMWMC | 0.0129 | 0.0641 | 0.0490 | 0.0838 | **0.6623** | 0.0594 | 0.0684 |
| | Steady states | 0.0189 | 0.0847 | 0.0782 | 0.0846 | 0.6649 | 0.0154 | 0.0532 |

## Data Availability

The data used to support the findings of this study are available from the corresponding author upon request.

## Ethical Approval

The manuscript is prepared in accordance with the ethical standards of the responsible committee on human experimentation and with the latest (2008) version of Helsinki Declaration of 1975.

## Conflicts of Interest

The authors declare that there are no conflicts of interest regarding the publication of this paper.

## Acknowledgments

The authors are very grateful to the Deanship of Scientific Research at King Khalid University, Kingdom of Saudi Arabia, for their administrative and technical support.

## References

[1] G. F. White, *Natural Hazards, Local, National, Global*, Oxford University Press, Oxford, UK, 1974.

[2] B. Edwards, M. Gray, and B. Hunter, "A sunburnt country: the economic and financial impact of drought on rural and regional families in Australia in an era of climate change," *Australian Journal of Labour Economics*, vol. 12, no. 1, p. 109, 2009.

[3] T. Oki and S. Kanae, "Global hydrological cycles and world water resources," *Science*, vol. 313, no. 5790, pp. 1068–1072, 2006.

[4] M. Svoboda and B. Fuchs, *Handbook of Drought Indicators and Indices*, World Meteorological Organization, Geneva, Switzerland, 2016.

[5] U. G. Bacanli, M. Firat, and F. Dikbas, "Adaptive neuro-fuzzy inference system for drought forecasting," *Stochastic Environmental Research and Risk Assessment*, vol. 23, no. 8, pp. 1143–1154, 2009.

[6] E. E. Moreira, C. A. Coelho, A. A. Paulo, L. S. Pereira, and J. T. Mexia, "SPI-based drought category prediction using loglinear models," *Journal of Hydrology*, vol. 354, no. 1-4, pp. 116–130, 2008.

[7] K. Lange, *Numerical Analysis for Statisticians*, Springer Science and Business Media, Berlin, Germany, 2010.

[8] C. Chatfield, *The Analysis of Time Series: an Introduction*, CRC Press, Boca Raton, FL, USA, 2016.

[9] M. M. Bateni, J. Behmanesh, J. Bazrafshan, H. Rezaie, and C. De Michele, "Simple short-term probabilistic drought prediction using Mediterranean teleconnection information," *Water Resources Management*, vol. 32, no. 13, pp. 4345–4358, 2018.

[10] J. Chang, Y. Li, Y. Wang, and M. Yuan, "Copula-based drought risk assessment combined with an integrated index in the Wei river basin, China," *Journal of Hydrology*, vol. 540, pp. 824–834, 2016.

[11] M. Rezaeianzadeh, A. Stein, and J. P. Cox, "Drought forecasting using Markov chain model and artificial neural networks," *Water resources management*, vol. 30, no. 7, pp. 2245–2259, 2016.

[12] X. Zhou, Y. Wang, and X. Zhou, "Precipitation estimation based on weighted Markov chain model," in *Proceedings of 2017 Seventh International Conference on Information Science and Technology (ICIST)*, pp. 64–68, IEEE, Da Nang, Vietnam, April 2017.

[13] P. Jun and W. Hao-han, "The application of weighted Markov chain in Hunhe flood control planning," *Jilin Water Resources*, vol. 8, p. 012, 2015.

[14] J. Chen and Y. Yang, "SPI-based regional drought prediction using weighted Markov chain model," *Research Journal of Applied Sciences, Engineering and Technology*, vol. 4, no. 21, pp. 4293–4298, 2012.

[15] X. C. Jiang and S. F. Chen, "Application of weighted Markov SCGM (1, 1) C model to predict drought crop area," *Systems Engineering-Theory and Practice*, vol. 29, no. 9, pp. 179–185, 2009.

[16] T. Zhang, J. Li, R. Hu, Y. Wang, and P. Feng, "Drought class transition analysis through different models: a case study in North China," *Water Science and Technology: Water Supply*, vol. 17, no. 1, pp. 138–150, 2017.

[17] P. J. Bhakta, *Markov chains for weighted lattice structures*, Ph. D. thesis, Georgia Institute of Technology, Atlanta, GA, USA, 2016.

[18] S. M. Dhanasekaran, T. R. Barrette, D. Ghosh et al., "Delineation of prognostic biomarkers in prostate cancer," *Nature*, vol. 412, no. 6849, pp. 822–826, 2001.

[19] D. Singh, P. G. Febbo, K. Ross et al., "Gene expression correlates of clinical prostate cancer behavior," *Cancer Cell*, vol. 1, no. 2, pp. 203–209, 2002.

[20] L. True, I. Coleman, S. Hawley et al., "A molecular correlate to the Gleason grading system for prostate adenocarcinoma," *Proceedings of the National Academy of Sciences*, vol. 103, no. 29, pp. 10991–10996, 2006.

[21] J. B. Welsh, L. M. Sapinoso, A. I. Su et al., "Analysis of gene expression identifies candidate markers and pharmacological targets in prostate cancer," *Cancer Research*, vol. 61, no. 16, pp. 5974–5978, 2001.

[22] Q. X. Zhou, Y. D. Wu, H. X. Fan, X. Wang, L. H. Sun, and S. Z. Wang, "Weight calculation and application of weighted Markov," *Journal of Harbin University of Commerce*, vol. 6, p. 028, 2014.

[23] D. Sengupta, U. Maulik, and S. Bandyopadhyay, "Weighted Markov chain based aggregation of biomolecule orderings," *IEEE/ACM Transactions on Computational Biology and Bioinformatics*, vol. 9, no. 3, pp. 924–933, 2012.

[24] T. B. McKee, N. J. Doesken, and J. Kleist, "The relationship of drought frequency and duration to time scales," in *Proceedings of the 8th Conference on Applied Climatology*, vol. 17, no. 22, pp. 179–183, American Meteorological Society, Boston, MA, USA, January 1993.

[25] S. M. Vicente-Serrano, S. Beguería, and J. I. López-Moreno, "A multiscalar drought index sensitive to global warming: the standardized precipitation evapotranspiration index," *Journal of Climate*, vol. 23, no. 7, pp. 1696–1718, 2010.

[26] Z. Ali, I. Hussain, M. Faisal et al., "A novel multi-scalar drought index for monitoring drought: the standardized precipitation temperature index," *Water Resources Management*, vol. 31, no. 15, pp. 4957–4969, 2017a.

[27] Z. Ali, I. Hussain, M. Faisal et al., "Forecasting drought using multilayer perceptron artificial neural network model," *Advances in Meteorology*, vol. 2017, Article ID 5681308, 9 pages, 2017.

[28] K. Takahashi, K. Morikawa, D. Takeda, and A. Mizuno, "Inventory control for a MARKOVIAN remanufacturing system with stochastic decomposition process," *International Journal of Production Economics*, vol. 108, no. 1-2, pp. 416–425, 2007.

[29] H. Y. Lee and S. L. Chen, "Why use Markov-switching models in exchange rate prediction?," *Economic Modelling*, vol. 23, no. 4, pp. 662–668, 2006.

[30] D. T. Crommelin and E. Vanden-Eijnden, "Fitting timeseries by continuous-time Markov chains: a quadratic programming

approach," *Journal of Computational Physics*, vol. 217, no. 2, pp. 782–805, 2006.

[31] Y. Gui and J. Shao, "Prediction of precipitation based on weighted Markov chain in Dangshan," in *Proceedings of the International Conference on High Performance Compilation, Computing and Communications*, pp. 81–85, ACM, Kuala Lumpur, Malaysia, March 2017.

[32] A. A. Paulo and L. S. Pereira, "Prediction of SPI drought class transitions using Markov chains," *Water Resources Management*, vol. 21, no. 10, pp. 1813–1827, 2007.

[33] A. K. Mishra, V. P. Singh, and V. R. Desai, "Drought characterization: a probabilistic approach," *Stochastic Environmental Research and Risk Assessment*, vol. 23, no. 1, pp. 41–55, 2009.

[34] K. Shatanawi, M. Rahbeh, and M. Shatanawi, "Characterizing, monitoring and forecasting of drought in Jordan River Basin," *Journal of Water Resource and Protection*, vol. 5, no. 12, pp. 1192–1202, 2013.

[35] J. Cohen, "A coefficient of agreement for nominal scales," *Educational and Psychological Measurement*, vol. 20, no. 1, pp. 37–46, 1960.

[36] J. Cohen, "Weighted Kappa: nominal scale agreement provision for scaled disagreement or partial credit," *Psychological bulletin*, vol. 70, no. 4, pp. 213–220, 1968.

[37] W. Z. Robieson, *On weighted Kappa and concordance correlation coefficient*, Ph.D. thesis, Graduate College, University of Illinois at Chicago, Chicago, IL, USA, 1999.

[38] T. S. King and V. M. Chinchilli, "A generalized concordance correlation coefficient for continuous and categorical data," *Statistics in Medicine*, vol. 20, no. 14, pp. 2131–2147, 2001.

[39] M. Gramer, J. Lemon, I. Fellows, and P. Singh, *Various Coefficients of Interrater Reliability and Agreement*, 2012.

[40] Q. Z. Chaudhry, "Construction of all Pakistan monsoon rainfall series 1901–2008," *Pakistan Journal of Meteorology*, vol. 6, no. 12, 2009.

[41] J. A. Khan, *The Climate of Pakistan*, Rehbar Publishers, Karachi, Pakistan, 1993.

[42] S. A. Awan, "The climate and flood risk potential of northern areas of Pakistan," *Science Vision*, vol. 7, no. 3–4, pp. 100–109, 2002.

[43] D. R. Archer and H. J. Fowler, "Spatial and temporal variations in precipitation in the Upper Indus Basin, global teleconnections and hydrological implications," *Hydrology and Earth System Sciences Discussions*, vol. 8, no. 1, pp. 47–61, 2004.

[44] Z. Ahmad, M. Hafeez, and I. Ahmad, "Hydrology of mountainous areas in the upper Indus Basin, Northern Pakistan with the perspective of climate change," *Environmental Monitoring and Assessment*, vol. 184, no. 9, pp. 5255–5274, 2012.

[45] J. H. Stagge, L. M. Tallaksen, L. Gudmundsson, A. F. Van Loon, and K. Stahl, "Candidate distributions for climatological drought indices (SPI and SPEI)," *International Journal of Climatology*, vol. 35, no. 13, pp. 4027–4040, 2015.

[46] K. Schittkowski, "EASY-FIT: a software system for data fitting in dynamical systems," *Structural and Multidisciplinary Optimization*, vol. 23, no. 2, pp. 153–169, 2002.

[47] J. R. Hosking, *L-Moments*, Wiley StatsRef: Statistics Reference Online, 2009.

[48] G. A. Spedicato and M. Signorelli, *The R Package "Markovchain": Easily Handling Discrete Markov Chains in R*, 2014.

[49] C. Zhang, X. Chen, X. Chen, K. Yu, G. Pan, and X. Zhang, "Analysis and prediction of annual precipitation based on weighted Markov chain in typical region of Taihu lake basin," *Bulletin of Soil and Water Conservation*, vol. 1, p. 032, 2015.

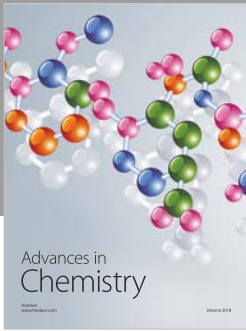
Advances in
Chemistry

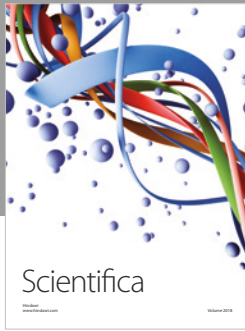
Scientifica

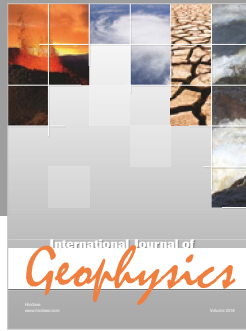
International Journal of
Geophysics

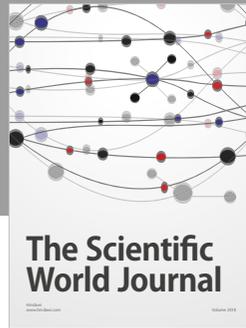
The Scientific
World Journal

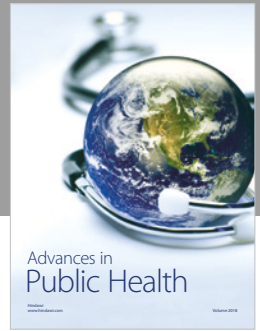
Advances in
Public Health

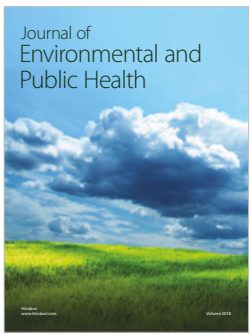
Journal of
Environmental and
Public Health

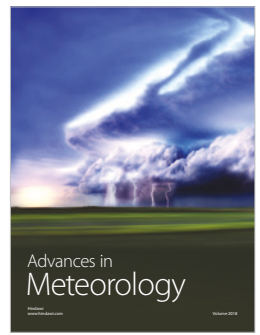
Advances in
Meteorology

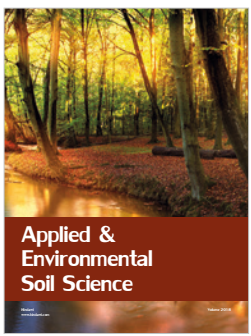
Applied &
Environmental
Soil Science

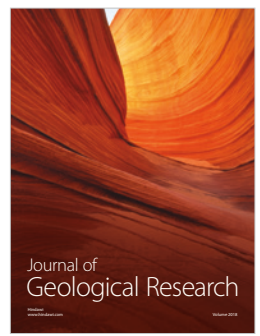
Journal of
Geological Research

Hindawi

Submit your manuscripts at
www.hindawi.com

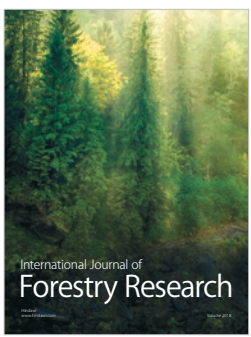
International Journal of
Forestry Research

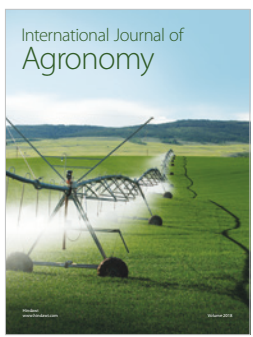
International Journal of
Agronomy

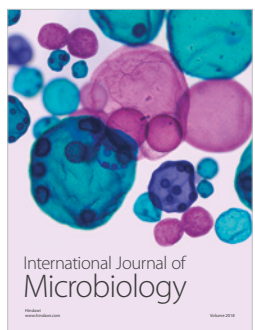
International Journal of
Microbiology

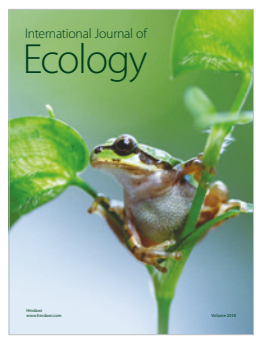
International Journal of
Ecology

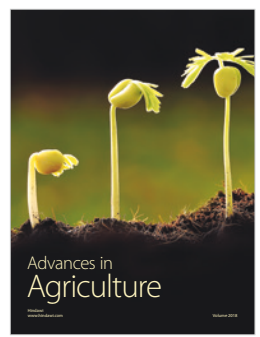
Advances in
Agriculture

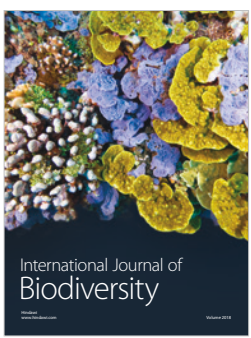
International Journal of
Biodiversity

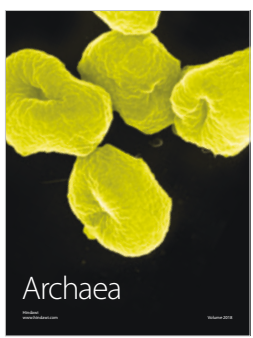
Archaea

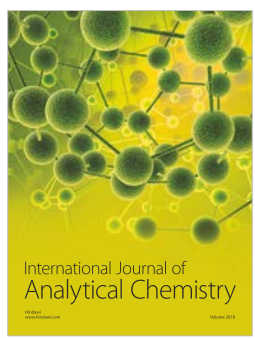
International Journal of
Analytical Chemistry

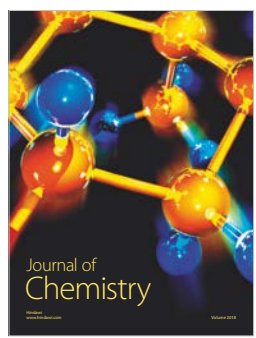
Journal of
Chemistry

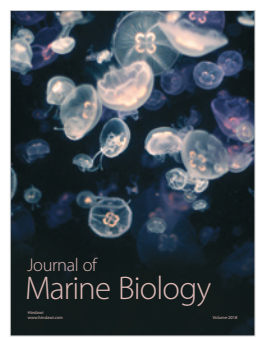
Journal of
Marine Biology